\documentclass[5p,twocolumn,number,sort&compress,times]{elsarticle}

\usepackage{amsmath}
\usepackage{amssymb}
\usepackage{graphicx}
\usepackage{bm}
\usepackage{bbm}

\usepackage{color}
\usepackage{verbatim}

\usepackage{hyperref}

\newcommand  {\nn}[1]        {\left\langle #1 \right\rangle}

\renewcommand{\vec}[1]       {\mathbf{#1}}                        
\newcommand  {\vr}           {\vec{r}}

\newcommand  {\zz}           {\mathbb{Z}}
\newcommand  {\defeq}        {\mathrel{\mathop:}=}

\begin{document}

\title{Topological Floquet states on a M\"obius band irradiated by circularly polarised light}

\author[itfuu]{A. Quelle}
\author[mpipks]{W. Beugeling}
\author[itfuu]{C. Morais Smith}

\address[itfuu]{Institute for Theoretical Physics, EMME$\Phi$, Utrecht University, Leuvenlaan 4, 3584 CE Utrecht, The Netherlands}
\address[mpipks]{Max-Planck-Institut f\"ur Physik komplexer Systeme, N\"othnitzer Stra\ss e 38, 01187 Dresden, Germany}

\date{\today}

\begin{abstract}
Topological states of matter in equilibrium, as well as out of equilibrium, have been thoroughly investigated during the last years in condensed-matter and cold-atom systems. However, the geometric topology of the studied samples is usually trivial, such as a ribbon or a  cylinder. In this paper, we consider a graphene M\"obius band irradiated with circularly polarised light.  Interestingly, due to the non-orientability of the M\"obius band, a homogeneous quantum Hall effect cannot exist in this system, but the quantum spin Hall effect can. To avoid this restriction, the irradiation is applied in a longitudinal-domain-wall configuration. In this way, the periodic time-dependent driving term tends to generate the quantum anomalous  Hall effect. On the other hand, due to the bent geometry of the M\"obius band, we expect a strong spin-orbit coupling, which may lead to quantum spin Hall-like topological states. Here, we investigate the competition between these two phenomena upon varying the amplitude and the frequency of the light, for a fixed value of the spin-orbit coupling strength. The topological properties are analysed by identifying the edge states in the Floquet spectrum at intermediate frequencies, when there are resonances between the light frequency and the energy difference between the conduction and valence bands of the graphene system. 
\end{abstract}

\maketitle


\section{Introduction}

The quantum Hall (QH), or Chern insulator is the first unveiled example of a topological insulator \cite{Luttinger1951,VonKlitzing1980}. With the discovery of the quantum spin Hall (QSH) insulator as a second example, the field of topological insulators was born \cite{Kane2005QSHE,Kane2005Z2,Bernevig2006,Konig2007}. Topological insulators are materials that insulate in their bulk, but have metallic edges. The possible phases of a system depend strongly on the symmetries it exhibits \cite{Kitaev2009,Ryu2010, Chiu2013}. Breaking a certain symmetry destroys the protection of the associated topological phase, allowing perturbations of the system to drive a phase transition. The QH and QSH insulators provide a telling example. The QSH insulator is protected by time-reversal symmetry (TRS): in the non-trivial phase, the QSH insulator hosts pairs of edge states that are conjugated by time reversal. A single time-reversal invariant pair (TRIP) of edge states cannot be gapped out by perturbations preserving the TRS, because for this process this symmetry would have to be broken. However, two TRIPs of edge states can be gapped out without breaking TRS. Thus, the parity of the number of TRIPs is conserved, giving a $\zz_2$-valued topological invariant for the system \cite{Kane2005QSHE,Kane2005Z2}. In contrast, the QH insulator is obtained when TRS is broken \cite{Haldane1988}, and any integer number of edge states is possible. This number of edge states is conserved, and yields a $\zz$-valued topological invariant \cite{Thouless1982}. The creation of a QH effect in a material can be achieved by applying a perpendicular magnetic field to a 2D sample \cite{VonKlitzing1980}. The creation of a QSH effect is more difficult, since it relies on strong spin-orbit (SO) coupling in the material. However, this has also been achieved in HgTe quantum wells \cite{Bernevig2006, Konig2007, Konig2008, Hasan2010}, Bi$_2$Se$_3$ and Bi$_2$Te$_3$ crystals \cite{Zhang2010} and Sn sheets \cite{Xu2013}. Furthermore, heavily bent graphene sheets exhibit significant SO coupling \cite{Steele2013}.

Aside from this, it is also possible to drive topological phase transitions in periodically driven out-of-equilibrium systems; one speaks then of Floquet topological insulators. Known examples are HgTe quantum wells \cite{Bernevig2006, Lindner2011, Katan2013}, as well as honeycomb lattices irradiated with circularly polarised light \cite{Inoue2010, Gu2011, Wang2013,Ezawa2013, Gomez-Leon2014}. The circularly polarised light breaks TRS and causes the appearance of a QH effect. This occurs because the rotation of the light field yields an effective dynamical magnetic field to the electrons \cite{Quelle2014}, which is known to create QH states \cite{Haldane1988}. Due to their out-of-equilibrium nature, these systems behave in a more complex fashion than their equilibrium counterparts. Nevertheless, the behaviour of Floquet systems in the absence of symmetry is well understood \cite{Kitagawa2010, Lindner2011, Katan2013, Rudner2013}, and important progress is being made in the investigation of systems with TRS \cite{Carpentier2014}.

The classification of topological phases, either in equilibrium or in the Floquet case, has been done for trivial sample geometries: the cylinder or the ribbon of infinite length. It is of fundamental interest to understand how the sample geometry influences the possible topological phases of a system. Preliminary investigations into this topic have been conducted by considering the M\"obius band, which unlike the cylinder and the ribbon, has only a single edge.
Actually, M\"obius bands have been investigated in a much wider context than just topological insulators. For example, an embedding of the band in three-dimensional space that minimises the deformation of the surface has been found through numerical calculation \cite{StarostinVanDerHeijden2007}. It has also been observed that the twist in the band can be considered to induce a gauge field in the Hamiltonian \cite{ZhaoEA2009}. In the case of graphene, the electronic \cite{YamashiroEA2004,JiangDai2008,CaetanoEA2009,Guo2009,WangEA2010,KorhonenKoskinen2014}, magnetic \cite{WakabayashiHarigaya2003,CaetanoEA2008}, and thermal \cite{JiangEA2010} properties of M\"obius bands have been investigated theoretically. Similar studies have been performed for boron nitride ribbons \cite{AzevedoEA2012}.  The experimental realisation of M\"obius bands in the lab has proven difficult as of yet, but has nevertheless already been achieved, with the fabrication of M\"obius bands made of NbSe$_3$ \cite{TandaEA2002}.

In the specific context of topological insulators, the behaviour of edge states on a M\"obius strip is known for equilibrium systems \cite{Huang2011,Beugeling2014}. A QSH effect is possible on the strip, just as on a ribbon. The SO Hamiltonian \cite{Kane2005Z2,Kane2005QSHE} is compatible with the twist of the M\"obius band and therefore well defined. More intuitively, the QSH effect can exist with only a single edge because each edge has counter-propagating edge states. In contrast, in the QH effect there exist counter-propagating edge states on opposite edges, a situation that is impossible with only a single edge. This impossibility can also be understood from the pseudovectorial nature of the magnetic field $\vec{B}$. Since $\vec{B}$ is a pseudovector, it changes sign under a change of unit normal. Consequently, a homogeneous magnetic field cannot be applied on the M\"obius strip, since the M\"obius band does not allow a continuous choice of unit normal. Preserving translational symmetry along the $x$-axis, it is possible to have perpendicular magnetic fields $\vec{B}=B(y)\hat{z}$ obeying $B(-y)=-B(y)$. By taking $B=B_0 \operatorname{sign}(y)$, one creates a QH-like system on the M\"obius strip, with the domain wall functioning as one of the two edges \cite{Beugeling2014}.


\begin{figure}[b]
\includegraphics{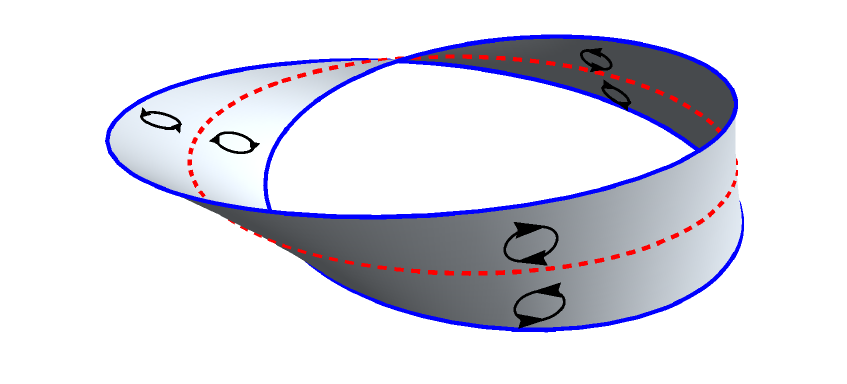}
\caption{The M\"obius band with a longitudinal domain wall (dotted curve) embedded in three-dimensional space. The circular arrows indicate the sense of the polarisation of the circularly polarised light.}\label{fig_mobius}
\end{figure}


In this paper, we consider a graphene M\"obius band irradiated by circularly polarised light. In this way, we expand the body of knowledge on topological states in a M\"obius geometry to the non-equilibrium regime. For our investigation, we write down a nearest-neighbour (NN) tight-binding Hamiltonian on the M\"obius strip, where the light is taken into account by incorporating a vector potential through the Peierls substitution. The light is included together with a longitudinal domain wall (see Fig.~\ref{fig_mobius}), since homogeneous irradiation is impossible, just as for a perpendicular $\vec{B}$ field. Since it is expected that graphene M\"obius bands also exhibit significant SO coupling due to the curvature of the surface, in analogy with carbon nanotubes \cite{Steele2013}, this term is also included in the model. These two terms in the Hamiltonian are diagonalised numerically for different values of light amplitude, light frequency and SO coupling strength, to demonstrate their comparative effects on the M\"obius band. We find that the circularly polarised light drives a QH-like state, where edge states propagate in one direction, while there are counter-propagating states localised at the domain wall. In the presence of SO coupling, a competition between the light and the SO coupling is observed, analogous to the case of a cylindrical geometry \cite{Quelle2014}, Our main result pertains the role played by the domain wall. For weak light intensities, the SO coupling dominates, and a ``weak QSH effect'' is seen, where helical edge states persist even in the presence of TRS breaking \cite{Yang2011,Goldman2012,Beugeling2012}, and no states occur at the domain wall. Once the light dominates, the QH-like state appears, together with the associated domain wall states. This holds true only for the gap at zero energy ($\varepsilon=0$). In contrast, in the gap at half the value of the light frequency, $\varepsilon=\omega/2$, which originates from photon resonances, the QH states always dominate. Furthermore, depending on the frequency $\omega$, two-photon resonances may create further edge states in the gap at $\varepsilon=0$ \cite{Quelle2014}. These edge states are always of a QH nature, and in the regime where light dominates, they counter-propagate with the QH states at the Dirac points. In this case, the $\varepsilon=0$ gap is topologically trivial, and the gapless edge states are no longer protected. This shows similarities with the weak QSH state, except that the propagation direction of the edge states is no longer coupled to spin, the counter-moving edge states occur at different momenta, and are driven by different terms (an effective magnetic field and photon resonances). This is in contrast to the crystalline insulators \cite{Fu2011,Slager2013}, which are also topologically trivial, but where the edge states are protected by a crystal symmetry.

The outline of the paper is as follows. In Section \ref{sect_theory}, we review the basic concepts used in this work, namely, the M\"obius band, and what kind of Hamiltonians are well-defined on this object, as well as Floquet theory. In Section \ref{sect_results}, we present our results. These include both the Hamiltonians that we use to model the irradiated M\"obius band, and the corresponding dispersion relations. First, we review the case of an irradiated cylinder, which we then compare to the irradiated M\"obius band without SO coupling. The differences are discussed, with an emphasis on their relation to the non-trivial geometry of the band. The M\"obius band with SO coupling is treated last, and compared with the band lacking the SO coupling. In this manner, the effect of the SO coupling can be distilled from the topological effects originating in the M\"obius geometry. In Section \ref{sect_conclusion}, our conclusions are presented.


\section{Theoretical prerequisites}%
\label{sect_theory}%

\subsection{M\"obius geometry}
The main geometrical feature of the M\"obius band is its non-orientability \cite{Frankel2004}, which results in the presence of a single edge, rather than two. This non-orientability means that no orientation, or continuous unit normal, can be chosen for the surface. Since the curvature $dA$ of the Berry connection is a pseudovector, it changes sign under a change of orientation, and hence the Chern number cannot be defined for non-orientable surfaces. For a M\"obius strip, one can also see what happens physically: the QH effect hosts counter-propagating states on opposite edges, but a M\"obius strip has only one edge, so that this state cannot exist here. On the other hand, a QSH state is possible on a M\"obius strip \cite{Huang2011}.

\begin{figure}
\includegraphics[width=\linewidth]{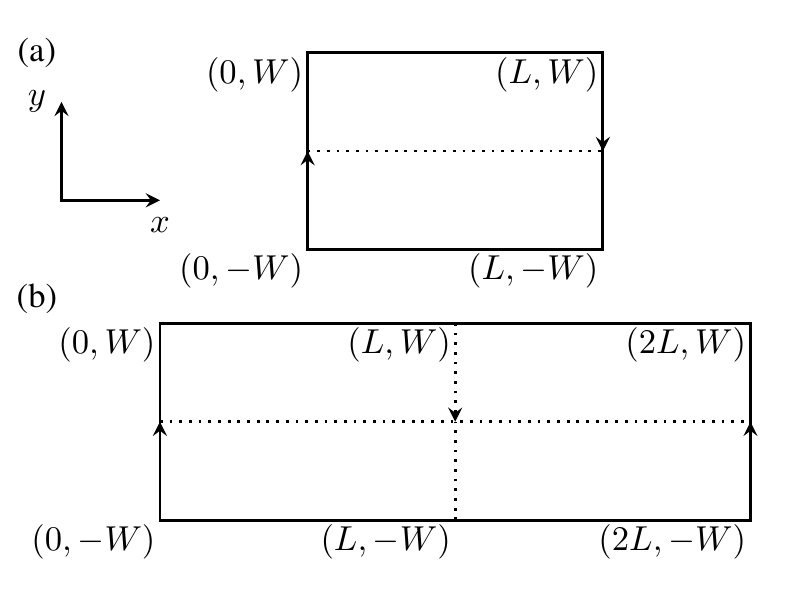}
\caption{(a) The M\"obius strip of circumference $L$ and width $2W$, shown as a rectangle of length $L$ with vertical edges identified. The identification is such that the arrows point in the same direction, inducing the characteristic twist in the surface. The horizontal dotted line shows the center of the strip, which is topologically a circle.
\noindent
(b) The double cover of the M\"obius strip, shown as a rectangle of length $2L$ and width $2W$. Applying the identification under $R$ for Eq.~\eqref{glidereflection}, the right half of the double cover is identified with the left half after mirroring in the $x$-axis. In this way the M\"obius band is obtained, and it lies twice in the double cover.}\label{fig_doublecover}
\end{figure}

One can bring the hard mathematics of topological classification to bear by viewing the M\"obius strip as a quotient space of the cylinder \cite{Beugeling2014} (see Fig.~\ref{fig_doublecover}). The orientable double cover of the M\"obius strip with circumference $L$ is topologically a cylinder with circumference $2L$ (as noted in Ref.~\cite{Beugeling2014}, the double cover has a $4\pi$ twist, but this feature is topologically unimportant). One then obtains the M\"obius strip from this double cover by identifying points that differ by a glide reflection
\begin{equation}\label{glidereflection}
R:(x,y)\mapsto (x+L,-y).
\end{equation}
Any Hamiltonian $H_M$ on the M\"obius strip can be extended uniquely to a Hamiltonian $H_C$ on the cylinder by requiring that it commutes with $R$. Vice versa, any Hamiltonian $H_C$ on the cylinder that commutes with $R$ descends to a Hamiltonian $H_M$ on the M\"obius strip. Furthermore, any wavefunction $\phi_M$ on the M\"obius strip can be uniquely extended to either an even or odd eigenfunction $\phi_C$ of $R$, and vice versa. The dispersion of $H_M$ is, therefore, precisely that of $H_C$ with half the degrees of freedom; one keeps the even eigenstates of $R$. This is possible because $[H_C,R]=0,$ so if $\phi$ is an eigenfunction of $H_C$, so are $\phi+R\phi$ and $\phi-R\phi$, which are even and odd since $R^2=1$. The topological classification of cylindrical systems with symmetries such as $R$ has been done in Ref.~\cite{Chiu2013}. As expected, one finds a $\zz_2$ topological invariant for a time-reversal invariant system, which corresponds to the QSH phase, and no topological behaviour in the absence of symmetries aside from $R$ \cite{Beugeling2014}. This last fact rigorously proves the impossibility of a QH phase on a M\"obius strip.


\subsection{Floquet theory}

Let us now investigate the effect of irradiating a graphene M\"obius  band with circularly polarised light. Since the latter is periodic in time, the Hamiltonian obeys $H(t+T)=H(t)$ for some period $T$. Hamiltonians of this form are analysed using Floquet theory \cite{Sambe1972}. Since the Hamiltonian is time-dependent, one does not consider the energy of the system because it is no longer a conserved quantity. As a substitute for the Hamiltonian, one studies the spectrum of the Floquet operator $U(T,0)$, where $U$ is the propagator given by
\begin{equation*}
U(t,t')=\mathcal{T}\left(\exp\left[-i\int_{t'}^t H(\tau)d\tau \right] \right),
\end{equation*}
where $\mathcal{T}$ denotes time ordering. 
Throughout the paper, we choose $\hbar=1$ for simplicity. Specifically, under fairly general conditions, $U(T,0)$ has a spectrum of eigenfunctions, which are quasi-periodic in time by the identity $U((n+1)T,nT)=U(T,0)$. The eigenfunctions of the Floquet operator are analogues of the bound states in an equilibrium system. Although they depend on time, their time-dependence is usually well controlled due to the quasi-periodicity. The Floquet operator has the defect of being unitary, rather than Hermitian, so one studies it indirectly through the Floquet Hamiltonian
\begin{equation}\label{floqham}
H_F \defeq -\frac{i}{T}\ln\left[U(T,0)\right].
\end{equation}
Here we choose $\ln$ to be the multi-valued logarithm, to emphasise the fact that the spectrum of $U(T,0)$ is periodic. Note that the physical content of the Floquet Hamiltonian lies in the identity $U(T,0)=\exp\left(iTH_F\right)$, where the right-hand side is again single-valued. The Floquet Hamiltonian has the interpretation of a time-averaged Hamiltonian. In the high-frequency limit, this Hamiltonian is the effective time-independent Hamiltonian that one obtains by averaging out the fast oscillation \cite{Hemmerich2010,Eckardt2005,Struck2012,Hauke2012}. For intermediate frequencies, resonances can appear, and the topological behaviour of the Floquet Hamiltonian can be more complicated than in the high-frequency limit \cite{Kitagawa2010,Lindner2011,Rudner2013}.


\section{Results: Dispersion and topological properties}%
\label{sect_results}

One prominent example of resonance-induced changes of topology is given by a graphene cylinder irradiated by circularly polarised light. The light breaks TRS in the sample, and causes a dynamically induced QH effect to appear \cite{Gomez-Leon2014, Usaj2014}. For frequencies of the circularly polarised light that allow resonances between the two energy bands, a gap is opened and chiral states appear. Furthermore, the chirality of the gaps is dependent on the frequency of the laser field. In samples where SO coupling is also present, the behaviour is even richer \cite{Quelle2014}. The circularly polarised light breaks TRS, and leads to a competition between the (weak) QSH and QH effect. As the intensity of the radiation is increased, a phase transition from the weak QSH to QH phase is observed. Here we show that on the M\"obius band, onto which homogeneous circularly polarised light cannot be applied unless a domain wall is introduced into the sample, this competition between QSH and QH effects leads to novel effects.

\subsection{Irradiated M\"obius graphene without SO coupling}


\begin{figure*}
\includegraphics[width=\linewidth]{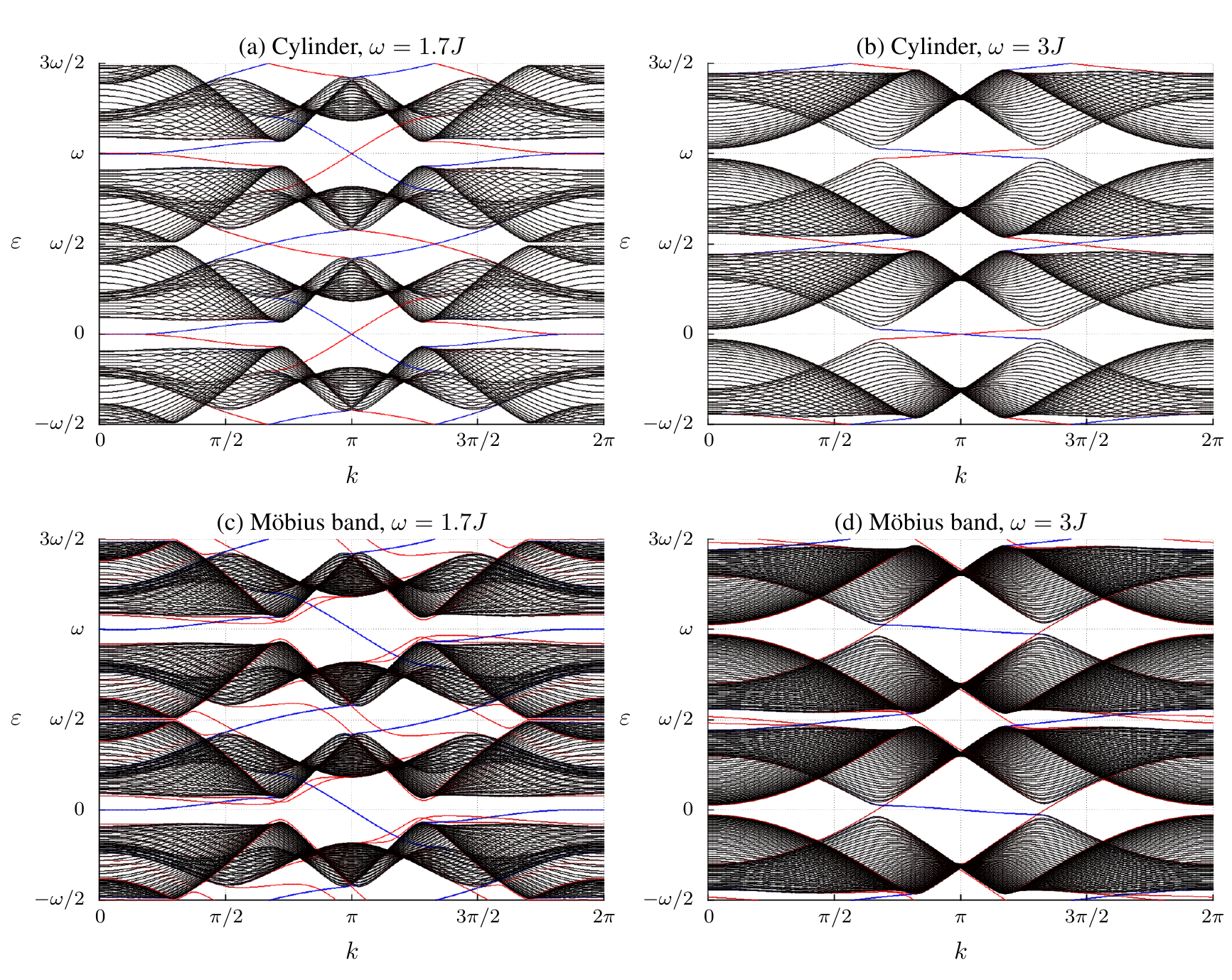}
\caption{The dispersion of $H_F$ given by Eqs.~\eqref{floqham} and \eqref{Hamiltonian}, for samples with zigzag edges and an even number of unit cells along the width of the ribbon. The light intensity is given by $E=J$. (a) Cylindrical geometry without domain wall, for $\omega=1.7J$. Edge states are coloured red/blue. (b) Same as (a), but for $\omega=3J$. (c) M\"obius geometry with domain wall, for $\omega=1.7J$. The edge states are in blue, while the states localised at the domain wall are in red. Hybridisation of the domain wall states lifts their degeneracy. At $\varepsilon=0$ the red states do not cross the gap; this is allowed because the chirality of this gap vanishes. (d) Same as (c), but for $\omega=3J$. }\label{fig_circlt}
\end{figure*}


To facilitate a comparison of the system on a M\"obius graphene band with the one on a cylinder, first the results on the cylinder are reviewed. This case is described by the tight-binding Hamiltonian
\begin{equation}\label{Hamiltonian}
\begin{split}
H_C    &\defeq  \sum_{\nn{i,j}}J_{ij}c^\dagger_i c_j\\
J_{ij} &\defeq  J\exp\left(-ie\int_{\vr_j}^{\vr_i}\vec{A}\cdot d\vec{s}\right). 
\end{split}
\end{equation}
Here, $\vec{A}$ is the vector potential incorporating the light, $J$ is the NN hopping parameter in graphene, and $\langle\cdot,\cdot\rangle$ denotes a sum over NN. Note that there is a suppressed spin index, since we are considering spinful electrons, but any spin interactions are neglected at this stage. In the sample plane $z=0$, circularly polarised light travelling in the $z$-direction is described by the vector potential 
\begin{equation}\label{vecpot}
\vec{A} \defeq
-\frac{E}{\omega}(\sin(\omega t),-\cos(\omega t)),
\end{equation}
in the Coulomb gauge with vanishing electrostatic potential. Furthermore, $E$ denotes the magnitude of the electric field. The Hamiltonian $H_C$ in Eq.~\eqref{Hamiltonian} defines a Floquet Hamiltonian $H_F$ according to Eq.~\eqref{floqham}. The dispersion of $H_F$ is plotted in Figs.~\ref{fig_circlt}(a) and (b), for two different values of frequency $\omega$. The spectrum is periodic, since $H_F$ is a multi-valued operator. In both cases, the spectrum exhibits two inequivalent energy gaps, one at $\varepsilon=0$ and one at $\varepsilon=\omega/2$. In the gap at $\varepsilon=0$, the light has given the Dirac electrons an effective mass, and the gap has opened in a topological way, creating a pair of edge states. In Fig.~\ref{fig_circlt}(a), an additional pair of edge states has appeared in this gap due to two-photon resonances ( $2\omega$) between the valence and the conduction band \cite{Quelle2014}. The two edge pairs have opposite chirality, so that the Hall conductivity vanishes. Because the counter-propagating states are separated in k-space, they do not automatically annihilate. However, these states are not protected by any symmetry, and may be annihilated by any interaction that couples the different $k$-values. This is in contrast to the so-called crystalline insulators, which may be in a trivial $\mathbb{Z}_2$ phase, and yet host edge states that are protected by symmetries of the crystal lattice \cite{Fu2011,Slager2013}. On the other hand, in Fig.~\ref{fig_circlt}(b), the photon frequency has increased, such that these two photon resonances are no longer possible, and only a single edge-state pair is seen. The gap at $\varepsilon=\omega/2$ is created by single photon resonances in both cases, and these resonances also create topological states. Due to the chiral nature of the circularly polarised light, all these states are of a QH nature. Changing the rotational direction of the light, by sending $t\mapsto -t$, or equivalently, $\omega\mapsto-\omega$, will reverse the propagation direction of all edge states by flipping the whole Floquet spectrum either vertically or horizontally.


\begin{figure}
\includegraphics[width=\linewidth]{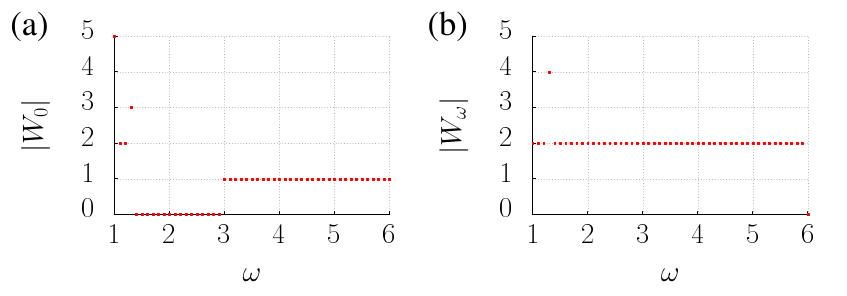}
\caption{The Floquet spectrum for irradiated graphene has two inequivalent band gaps, one at $\varepsilon=0$, and one at $\varepsilon=\omega/2$. The chiralities of the two gaps are plotted against $\omega$ for $E=J$: (a) The chirality $|W_0|$ of the gap at $\varepsilon=0$. (b) The chirality $|W_\omega|$ of the gap at $\varepsilon=\omega/2$.}\label{fig_phases}
\end{figure}


Under a change of orientation (i.e., of the direction of the unit normal to the surface), the Poynting vector of a coherent light wave changes sign. This shows that a M\"obius band cannot be homogeneously irradiated with circularly polarised light. Alternatively, $H_C$ does not commute with $R$, because $\vec{A}$ from Eq.~\eqref{vecpot} is not symmetric under reflection. However, by allowing a domain wall at the center of the strip (see the horizontal dotted line in Fig.~\ref{fig_doublecover}(a)), the inclusion of a laser field in the Hamiltonian is possible. In this way, the non-trivial topology of the M\"obius band enforces constraints on the system.

To incorporate this domain wall, we use the Hamiltonian $H_C$ on the double cover of the M\"obius strip (Fig.~\ref{fig_doublecover}(b)), but replace $\vec{A}$ by
\begin{equation}\label{Rvecpot}
\vec{A}\defeq\left\{
\begin{aligned}
-&\frac{E}{\omega}(\sin(\omega t),-\cos(\omega t))&y<0,\\
-&\frac{E}{\omega}(\sin(\omega t),\cos(\omega t))&y>0.\\
\end{aligned}\right.
\end{equation}
Now the $y$-component of the vector potential changes sign along the domain wall, which corresponds to a sign change in the Poynting vector of the light field. It is easily checked that $H_C$ given by Eqs.~\eqref{Hamiltonian} and \eqref{Rvecpot} commutes with the glide reflection $R$ from Eq.~\eqref{glidereflection}, and therefore it descends to a Hamiltonian $H_M$ on the M\"obius strip.

The Hamiltonian $H_M$ obtained from Eqs.~\eqref{Hamiltonian} and \eqref{Rvecpot} defines a Floquet Hamiltonian $H_F$ according to Eq.~\eqref{floqham}. In Figs.~\ref{fig_circlt}(c) and (d) the dispersion of $H_F$ is shown for different values of $\omega$. It can be seen that the chirality of the gaps, defined as the number of edge states taking into account their direction of motion, depends on the frequency of the applied light. The chiralities of these gaps are shown in Fig.~\ref{fig_phases} for $E=J$ and a range of frequencies $\omega/J$. In Fig.~\ref{fig_phases}(a) the chirality $W_0$ of the gap at $\varepsilon=0$ is shown; in Fig.~\ref{fig_phases}(b) the chirality $W_\omega$ of the gap at $\varepsilon=\omega/2$ is shown. From Fig.~\ref{fig_phases} it is apparent that between $\omega=1.7J$ and $\omega=3J$, $W_0$ has changed, while $W_\omega$ has not. This shows that $W_0$ and $W_\omega$ are independent, and motivates the choices of $\omega$ in Fig.~\ref{fig_circlt}.

Comparing Figs.~\ref{fig_circlt}(a) and (c) with (b) and (d), one sees that on the M\"obius strip a QH-like state is created, similar to that on the cylinder. Because the light is circularly polarised, reflection symmetry in the $x$-direction is broken. Consequently, there is a preferred direction of propagation on the edge, and it is clear that the blue edge states on the cylinder survive unaltered on the M\"obius strip. Changing the polarisation causes the propagation direction to reverse. Furthermore, the dispersion of the edge states (blue) is unaltered, because the edge states are localised and do not notice the non-trivial geometry of the strip. The difference between the M\"obius band and the cylinder can be seen in the states localised on the domain wall (the states in red). On the M\"obius band, the domain wall starts functioning as a second edge, analogous to the edge on the cylinder for which the edge states are coloured red in Figs.~\ref{fig_circlt}(a) and (b). However, the domain wall is not a hard boundary: hopping can occur across it. This causes the states localised on the domain wall to hybridise, splitting their degeneracy. Indeed, the red states are non-degenerate, while the blue states are doubly degenerate. The preservation of the degeneracy of the blue states can be understood from the lack of hybridisation between them, due to their spatial separation. In the interpretation of the two components of the double cover as two spin components, this degeneracy may be viewed as a spin degeneracy \cite{Beugeling2014}. The glide reflection, which is used to obtain a M\"obius band from a cylinder by such a glide reflection (see Fig.~\ref{fig_doublecover}), involves a mapping of one component onto the other, which may be interpreted as spin flip in this framework. The hopping across the domain wall connects the two (spin) components, which leads to the hybridisation and the consequent lifting of the degeneracy.
 
The equality of the number of states at the edge and at the domain wall is to be expected on physical grounds. The cylindrical system with a domain wall can be interpreted as two cylindrical systems interacting at their boundary. The hopping across this boundary causes hybridisation of the edge states. Tuning the hopping strength across the domain wall shows that this is indeed the case; the hybridisation weakens as the hopping across the domain wall becomes smaller. If this hopping is put to zero, the dispersion for the cylinder is recovered \cite{Beugeling2014}.

It is reasonable to suppose that the hybridisation of the domain wall states cannot affect the total chirality of each gap. This implies that the total chirality of the domain wall states is opposite to that of the edge states. Interestingly, the hybridisation does allow for a change in the number of crossing domain wall states, as long as the chirality is preserved. This can be seen in the gap around $\varepsilon=0$ in Fig.~\ref{fig_circlt}(c), where the two blue pairs of edge states are counter-propagating, and the four red states do not cross. This occurs because there is no symmetry protecting the counter-propagating edge states in the gap at $\varepsilon=0$. Due to the lack of topological protection, the hybridisation at the domain wall can annihilate the gapless states. 

Finally, note that the appearance of QH-like edge states on the irradiated M\"obius band does not contradict the impossibility of a QH effect on this system. The domain wall breaks translational symmetry along the $y$-direction, making this a local QH effect. Such a local QH effect can exist, because locally, the M\"obius band looks like a two-dimensional plane.

\subsection{Irradiated M\"obius graphene with SO coupling}


\begin{figure*}
\includegraphics[width=\linewidth]{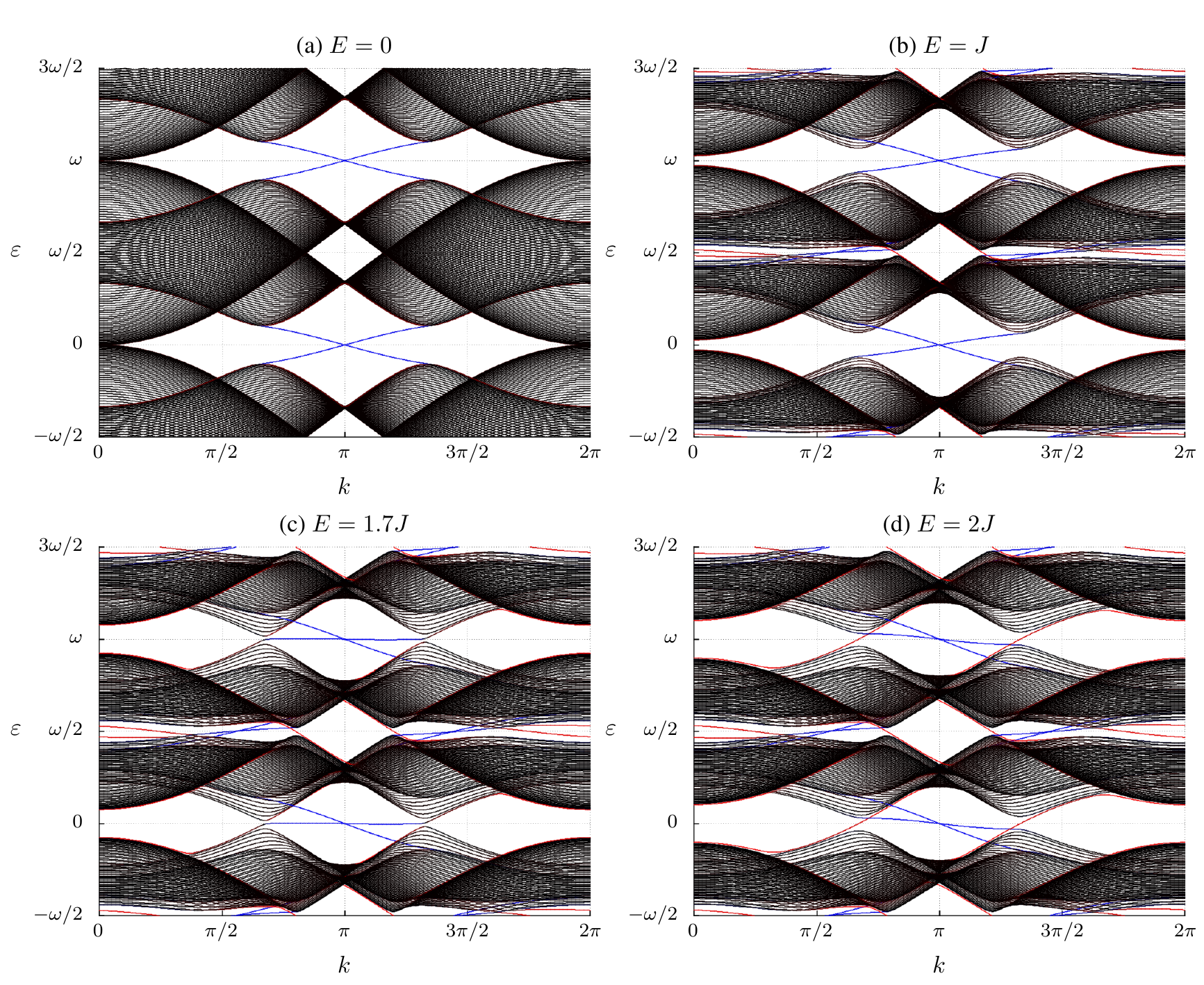}
\caption{The dispersion of $H_F$ given by Eqs.~\eqref{floqham} and \eqref{FloqHamISO} for M\"obius bands with zigzag edges and an even number of unit cells along the width of the band. The frequency of the light is $\omega=3J$, while the SO coupling strength is $\lambda=0.06J$. (a) In the absence of light ($E=0$), the ordinary QSH effect is observed. The spectrum has been periodically continued in $\varepsilon$ for ease of comparison. (b) For $E=J$, a weak QSH effect is seen in the gap at $\varepsilon=0$ since the SO coupling dominates. There, SO coupling also prevents the appearance of states at the domain wall. At the photon resonances, the QH-like states appear in the same manner as in the absence of SO coupling. (c) For $E=1.7J$, the gap closes at $\varepsilon=0$ since the SO coupling and the light annihilate each other. The phase transition occurs at this point. (d) For $E=2J$, the light dominates the SO coupling; QH-like states now occur at $\varepsilon=0$ as well as at $\varepsilon=\omega/2$. Apart from a lifting of the spin degeneracy of the edge states, the dispersion looks similar to that in Fig.~\ref{fig_circlt}(d).}\label{fig_circltso}
\end{figure*}


Although flat graphene samples have a negligible SO coupling \cite{Min2006}, recently it has been shown that in strongly bent graphene samples such as carbon nanotubes, SO coupling can take appreciable values \cite{Steele2013}. A graphene M\"obius band of small radius would fall into this category. For accuracy, the effect of SO coupling has to be included in the Hamiltonian. It is known that the interplay between circularly polarised light and SO coupling creates a competition between a QH and QSH phase in cylindrical systems \cite{Quelle2014}. This is a dynamical version  of the competition created by applying a perpendicular $\vec{B}$ field to a graphene sample with SO coupling \cite{Beugeling2014}. Such a dynamical competition creates effects not seen in the static case. We will show that these effects also appear on the M\"obius band, together with the hybridisation effects at the domain wall apparent in Fig.~\ref{fig_circlt}. The description of this behaviour constitutes the main result of this paper.

To model the system, we use the Hamiltonian \cite{Kane2005Z2,Kane2005QSHE,Quelle2014}
\begin{equation}\label{FloqHamISO}
H_C(t)=-\sum_{\langle i,j \rangle}J_{ij}(t) c^\dagger_i c_j-i\sum_{\langle\langle i,j \rangle\rangle}\lambda_{ij}(t)\nu_{ij}c^\dagger_i\hat{s}_z c_j,
\end{equation}
where double brackets denote a sum over next-nearest neighbours, $\hat{s}^z$ is a Pauli matrix, and $\nu_{ij}=\pm1$ where the sign depends on the cross product of the two NN vectors that connect sites $i$ and $j$ via the uniquely defined intermediate site. Furthermore,
\begin{equation}\label{Peierls}
J_{ij}(t)=J \exp\left[-i e \int_{\vr_j}^{\vr_i} \vec{A}(t)\cdot d\vec{s}\right],
\end{equation}
and 
\begin{equation}\label{PeierlsISO}
\lambda_{ij}(t)=\lambda \exp\left[-i e \int_{\vr_j}^{\vr_i} \vec{A}(t)\cdot d\vec{s}\right],
\end{equation}
with $\lambda$ denoting the ISO coupling strength. Once again, $\vec{A}$ is given by Eq.~\eqref{Rvecpot}. Because $H_C$ in Eq.~\eqref{FloqHamISO} is symmetric under the glide reflection $R$, it descends to a Hamiltonian $H_M$ on the M\"obius band. We use the $H_M$ obtained from Eq.~\eqref{FloqHamISO} to model the interaction between the SO coupling and the circularly polarised light on the M\"obius band. This is done by examining the spectrum of the corresponding Floquet Hamiltonian $H_F$.

In Fig.~\ref{fig_circltso}(a), the dispersion for a graphene M\"obius band without any radiation is observed. For ease of comparison with the irradiated case, the spectrum has been periodically continued in $\varepsilon$ with period $\omega=3J$. The gap at $\varepsilon=0$ has opened up and a pair of counter-propagating edge states has appeared on the single edge of the M\"obius band, in agreement with previous studies that showed the occurrence of a QSH effect in M\"obius graphene bands in the presence of SOC \cite{Huang2011, Beugeling2014}.

In Fig.~\ref{fig_circltso}(b), the dispersion is shown after the circularly polarised light has been turned on with $\omega=3J$ and $E=J$. Even though the circularly polarised light breaks TRS (time reversal changes the rotational direction of the light, and hence the propagation direction of the QH states), a QSH-like state is still observed. This state is called \emph{weak} QSH effect, because it is not protected by TRS, but this does not necessarily lead to the appearance of scattering effects that would destroy the QSH state \cite{Yang2011,Goldman2012,Beugeling2012}. Although the net magnetic field felt by the electrons due to the circularly polarised light in principle competes with the (weak) QSH state and tends to destroy it, at this intensity it just leads to unequal magnitude in the slope of the two edge states. Indeed, for one spin value the two effects enhance each other, while for the other the two effects oppose each other. Also note the absence of hybridised gapless states at the domain wall. The SO coupling only creates localised states on the edge of the M\"obius band, and it \emph{prevents} the appearance of states localised at the domain wall, as long as it dominates. Finally, the circularly polarised light also opens up a gap at $\varepsilon=\omega/2$, which is caused by photon resonances between the valence and conduction bands \cite{Quelle2014}. In this gap, QH-like states are observed, similar to those in Fig.~\ref{fig_circlt}(d). Tuning the strength of the SO coupling shows that in this gap only the QH-like states are possible. This is because the gap is opened by photon resonances, and the topological states are similarly created by these resonances. The SO coupling cannot create such resonances, and therefore cannot change the topological nature of this gap either. 

In Fig.~\ref{fig_circltso}(c) the intensity of the light is further increased. By increasing the light intensity, the edge state where QH and QSH effect oppose each other becomes increasingly dispersionless. At the light intensity $E\approx1.7J$, the two effects annihilate each other, and the gap at $\varepsilon=0$ closes. The appearance of a state localised at the domain wall is also visible, but it is not topological yet, since the circularly polarised light not yet dominates.

In Fig.~\ref{fig_circltso}(d) the light intensity is further increased, and now the effect of the circularly polarised light is dominant. In the gap at $\varepsilon=0$, the TRS is broken to such an extent that the QSH effect has vanished, and a QH-like state has appeared. Two edge states with fixed propagation direction appear and two counter-propagating states appear at the domain wall, in accord with Fig.~\ref{fig_circlt}(d). The SO coupling lifts the degeneracy of the (blue) edge states, but is not sufficiently strong to alter the QH nature of these edge states. The (red) states at the domain wall already have their degeneracy lifted by the hybridisation, as discussed before.

This competition between the SO coupling and the circularly polarised light is also seen for $\omega=1.7J$ (not shown). Here, the gapless state at $k=0$ in the $\varepsilon=0$ gap causes a mentionable effect. For intensities where the light dominates, similar behaviour to that in Fig.~\ref{fig_circlt}(c) is observed. Specifically, there are counter-propagating edge states, while no states cross at the domain wall. For frequencies where the SO coupling dominates, the chirality of the gap is non-zero, and there are no domain wall states at the Dirac points around $k=\pi$. In this case, there is a gapless domain-wall state at $k=0$; in the absence of SO coupling, this gapless state annihilates against those at the Dirac points, as can be seen in Fig.~\ref{fig_circlt}(c).


\section{Conclusion}\label{sect_conclusion}%
The application of circularly polarised light to a graphene M\"obius band induces QH-like states in the Floquet spectrum of the system similarly to the situation in graphene \cite{Gomez-Leon2014, Usaj2014}. Important differences arise in the dispersion due to the non-trivial topology of the M\"obius band. These differences are qualitatively similar to those for the time-independent QH states on the M\"obius strip discussed in Ref.~\cite{Beugeling2014}. Specifically, since the M\"obius band has only one edge, the function of second edge is taken over by the domain wall. The domain wall is, however, not a solid boundary, since electrons can hop across it. This causes a hybridisation of the domain wall states and lifts their degeneracy, whereas the states at the edge of the M\"obius band remain degenerate. If one takes into account SO coupling induced by strong curvature \cite{ZhaoEA2009}. on the M\"obius band, a competition between the SO coupling and the circularly polarised light is observed. Qualitatively, the observed behaviour is reminiscent of that on a carbon nanotube \cite{Quelle2014}. However, on the M\"obius strip, an important novelty occurs due to the presence of a domain wall. The SO coupling does not create topological states at the domain wall, and for parameter values where it dominates, it even destroys the ones created by light. As the intensity of the circularly polarised light increases, localised states slowly appear at the domain wall, and they become topological when the light starts dominating the behaviour of the gap.
For frequencies $\omega$ of the light where two photon resonances create a second pair of gapless states, this effect creates further novelties. The gapless states at the resonance are always of a QH nature, and hence there are domain wall states. They are gapless as long as the SO coupling destroys the domain wall states at the Dirac points, but when the light dominates, all the domain wall states annihilate each other. They can do this because the gap is trivial when the light dominates; when the SO coupling dominates, the gap is non-trivial.

This work has shed light on the impact the non-trivial M\"obius geometry has on Floquet topological insulators in graphene. We expect that these results help to understand the role of non-trivial geometries on the electronic structure of materials on a fundamental level, and to motivate further research into this topic.

\section*{Acknowledgements}
We thank M. Goerbig, V. Juri\v{c}i\'{c}, and A. Grushin for useful comments and discussions. The work by A.Q.\ is part of the D-ITP consortium, a program of the Netherlands Organisation for Scientific Research (NWO) that is funded by the Dutch Ministry of Education, Culture and Science (OCW). C.M.S.\ acknowledges NWO for funding within the framework of a VICI program.


%

\end{document}